\documentclass[aps,prl,superscriptaddress,reprint]{revtex4-1}
\usepackage{graphicx}
\usepackage{amsmath}
\usepackage{amssymb}
\usepackage{amsfonts}
\usepackage{filecontents}
\usepackage{color}
\usepackage[normalem]{ulem} 
\usepackage{soul}
\usepackage{epstopdf}
\usepackage{hyperref}
\definecolor{gold}{rgb}{0.85,.66,0}

\begin{document}

\title{Phonon-assisted population inversion of a single quantum dot}
\author{J. H. Quilter} 
\author{A. J. Brash}
\author{F. Liu} \email[email: ]{fengliu@sheffield.ac.uk}
\affiliation{Department of Physics and Astronomy, University of Sheffield, Sheffield, S3 7RH, United Kingdom}
\author{M. Gl\"{a}ssl}
\author{A. M. Barth}
\author{V. M. Axt}
\affiliation{Institut f\"{u}r Theoretische Physik III, Universit\"{a}t Bayreuth, 95440 Bayreuth, Germany}
\author{A. J. Ramsay}
\affiliation{Hitachi Cambridge Laboratory, Hitachi Europe Ltd., Cambridge CB3 0HE, United Kingdom}
\author{M. S. Skolnick}
\author{A. M. Fox}
\affiliation{Department of Physics and Astronomy, University of Sheffield, Sheffield, S3 7RH, United Kingdom}
\date{\today}
\begin{abstract}

We demonstrate an approach to realize
the population inversion of a single InGaAs/GaAs quantum dot, which is driven by
a laser pulse tuned within the neutral exciton phonon sideband. The inversion is
achieved by rapid thermalization of the optically dressed states via
phonon-assisted relaxation. A maximum exciton population of 0.67 $\pm$ 0.06 is
measured for a laser tuned 0.83~meV to higher energy and the phonon sideband is
mapped using a two-color pump-probe technique. Our experiments  
reveal that, in accordance with theory, the phonon-bath provides additional functionality for an optically driven quantum dot qubit.
\end{abstract}

\maketitle

It is a basic tenet of laser physics that a population inversion cannot be
achieved through incoherent excitation of a two-level atom. At best, a 
laser pulse with duration longer than the coherence time $T_2$ can only drive
the system to the transparency point where the populations of the upper and
lower levels are equal\cite{book:lasers}. However, if the two-level atom is coupled to
a vibrational continuum, it has been predicted that inversion can be possible
even in the incoherent regime through the interaction of the dressed states with
the Boson bath\cite{Legget1987}.  Excitons in semiconductor quantum dots (QDs)
form a near ideal system for investigating these effects, since their behavior
approximates well to that of a two-level atom\cite{Stievater2001}, while their
coupling to the acoustic phonons in the crystal provides a mechanism to
thermalize the dressed states.

The possibility of creating population inversion in QDs through
phonon coupling was first investigated for microwave-driven electrostatic quantum
dots\cite{Petta2004}. Recently, it has been demonstrated that the
conditions for population inversion can be met via a microwave Raman
effect\cite{Stace2005,Colless2014}. Theoretical work has indicated that similar
effects should be possible for optically-driven excitons
{\cite{Glassl2011, Reiter2012, Glassl2013, Hughes2013a}}. 
The underlying mechanism is the coupling of the
excitons to longitudinal acoustic (LA) phonons through the deformation
potential\cite{Krummheuer2002}, which generates sidebands in the excitonic
spectra \cite{besombes} that can also be observed in four-wave mixing\cite{Borri2005} and resonance
fluorescence experiments\cite{Weiler2012}, as well as through off-resonant
coupling of excitons to nano-cavities\cite{Hohenester2009,Madsen2014}. 
In a strong driving field regime evidence for phonon induced relaxation between
optically dressed states is observed in the intensity damping of Rabi rotations
\cite{Ramsay2010a,Ramsay2010,Monniello2013}, and more recently in adiabatic
rapid passage experiments \cite{chirp, Mathew2014}.


In this letter we report a population inversion of the excitonic two-level system
of a single InGaAs/GaAs QD under strong laser pumping at positive detuning with the driving laser to higher energy than the QD zero phonon transition. We compare the results to the theoretical
predictions of Ref.~\cite{Glassl2013}. The population
inversion is achieved in an incoherent regime where the dephasing time is
shorter than the laser pulse duration. Pump-probe measurements are presented, where the
phonon-assisted population inversion is observed clearly as a gain-like
dip in the photocurrent absorption spectrum. Furthermore, the dependence of the
exciton population on the pump detuning is measured. 
 The experiments are supplemented by simulations based on the path-integral
approach described in Ref.~\cite{Vagov2011}
and good agreement is obtained with the theory. 
This work proves experimentally that
 the phonon-bath can provide additional functionality for
an optically driven QD qubit, rather than simply acting as a source of
decoherence.
   


\begin{figure}
\includegraphics[width=8.6cm]{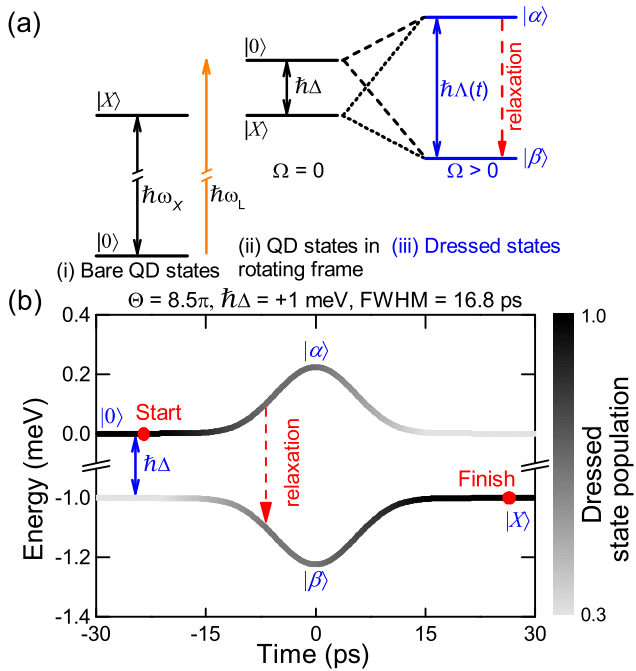}
\caption {Color online. (a) Bare QD states viewed in the (i) lab frame and (ii) rotating frame. $|0\rangle$ and $|X\rangle$ denote the crystal ground-state and exciton state. $\hbar\omega_{X}$ is the transition energy. $\omega_\text{L}$ is the angular frequency of the laser. $\hbar\Delta$ is the detuning from the exciton transition. (iii) Optically dressed states $|\alpha\rangle$ and $|\beta\rangle$. (b) Evolution of the dressed QD energy levels with time during the absorption of an 8.5$\pi$ pulse with $\hbar\Delta =$~+1~meV. The greyscale of the curves corresponds to the instantaneous population of each state.}
\label{fig:dressed_states}
\end{figure}

The physical mechanism for generating the population inversion can be understood as follows. A circularly-polarized laser pulse with a large pulse area excites the neutral exciton transition of an InGaAs/GaAs quantum dot at a positive detuning within the phonon-sideband, which typically peaks around 1~meV above the exciton (see Fig.~\ref{fig:detune_map}(a) insert). Since the laser bandwidth of 0.2~meV is large compared with the fine-structure splitting of $13~\mathrm{\mu eV}$, the dynamics of the exciton spin can be neglected\cite{Wang2005}. Also, since  the laser is far detuned from the two-photon biexciton transition, the QD can be treated as a two-level system composed of two bare states: a crystal ground-state $\vert 0\rangle$, and neutral exciton $\vert X\rangle$ (see Fig.~\ref{fig:dressed_states}(a)(i)), with respective populations: $C_0$ and $C_X$. With the presence of a laser pulse, the two bare states are optically dressed (see Fig.~\ref{fig:dressed_states}(a)(ii), (iii)). The Hamiltonian in a rotating frame reads:
\begin{equation}
H_{\mathrm{QD}}= -\hbar\Delta| X\rangle\langle X|+\frac{\hbar\Omega(t)}{2}| 0\rangle\langle X| + H.c.,
\end{equation}
where the detuning $\Delta = \omega_{\mathrm{L}} - \omega_X$ , and $\Omega(t)$ is the Rabi frequency, which varies in time following the envelope of the laser pulse. The energy eigenstates of $H_{\mathrm{QD}}$, $|\alpha\rangle,|\beta\rangle$ are described by an admixing angle $2 \theta(t) = \mathrm{atan}( \Omega(t) / \Delta)$, and are split by the effective Rabi energy $\hbar\Lambda(t)=\hbar\sqrt{\Delta^2+\Omega(t)^2}$. We define the bare pulse area $\Theta = \int_{-\infty}^{+\infty} \Omega (t) \,\mathrm{d}t$.

The QD resides in a crystal lattice, and the exciton interacts with the LA phonons due to the deformation potential \cite{Krummheuer2002}. 
In the absence of a laser-field, the exciton-phonon interaction leads to non-exponential pure-dephasing of the excitonic dipole, as observed in time-resolved four-wave mixing experiments \cite{Borri2005}. In the presence of a strong laser, due to the coupling between the excitonic component of the optically dressed states, there is relaxation between them, accompanied by the emission/absorption of a phonon with energy equal to the effective Rabi-splitting. This process gives rise to intensity damping of Rabi rotations when resonantly pumping
\cite{Ramsay2010a,Ramsay2010}, 
and also explains the difference in the population inversions created by adiabatic rapid passage when using laser pulses of 
positive and negative chirp \cite{chirp, Mathew2014}. 

Figure \ref{fig:dressed_states}(b) depicts the dynamics of the phonon-assisted
population inversion of the quantum dot exciton. Initially for positive detuning
($\Delta>0$), the crystal ground-state is the higher-energy dressed state. When
the laser is applied, the excitonic component of the higher-energy state $|\alpha \rangle$
increases, activating phonon-relaxation to the lower energy-level $|\beta \rangle$.  The
greyscale indicates the time-dependent populations of each dressed state, as
calculated by our path-integral method, and shows the
continuous transfer of population from $|\alpha \rangle$ to $|\beta \rangle$. As
the laser intensity drops off at the end of the pulse, the admixing of the
dressed states is reduced and the phonon relaxation is deactivated. If the
time-integrated phonon-relaxation is strong, the final occupation of the
lower-energy, exciton-like, dressed state can dominate and a population
inversion in the excitonic basis occurs. According to our theoretical model,
near-unity exciton population that is robust against variations in both
the pulse-area and the
detuning can be achieved at high pulse-areas \cite{Glassl2013}. The
amount of inversion is ultimately limited by the thermal occupation of the
two states to $C_X-C_0=\tanh ( \hbar\Delta / 2 k_{\mathrm{B}}T)$
\cite{Glassl2013}. In our experiments with $ \hbar\Delta = 0.83~\mathrm{meV}$
and $T=4.2~\mathrm{K}$, this implies a maximum exciton population of 0.91.


The experiments were performed on a device consisting of a
layer of InGaAs/GaAs quantum dots embedded in the intrinsic region of an
n-i-Schottky diode structure. The sample is held at 4.2~K in a helium bath
cryostat and is excited at normal incidence with circularly polarized Gaussian
pulses, derived by pulse-shaping of the output from a 100~fs Ti:sapphire laser
with repetition rate of 76.2~MHz. The FWHM of the electric field amplitude is
16.8~ps. Single QD peaks are observed in our sample at energies close to 1.3~eV.
Photocurrent detection is used, where photo-generated electron-hole pairs tunnel
from the quantum dot under an applied electric-field, leading to a change in
photocurrent proportional to the final occupation of the exciton or biexciton states. Full details of the wafer structure can be found in Ref.~\onlinecite{Kolodka2007}.

\begin{figure}
\includegraphics[width=8.6cm]{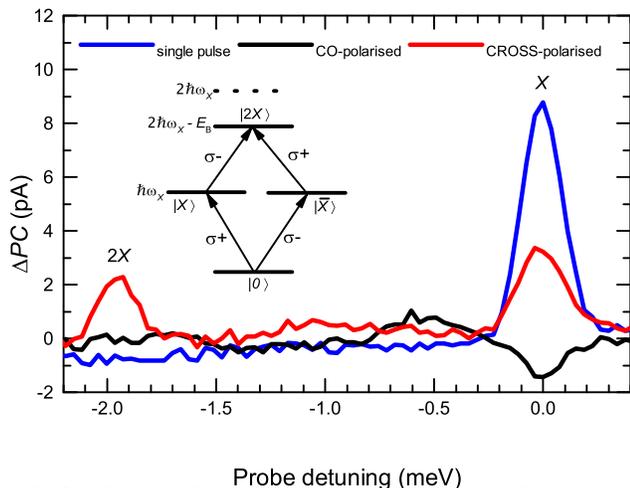}
\caption {Color online. Photocurrent signal $\Delta PC$ as a function of the probe detuning. (Blue) A single probe-pulse only spectrum is presented for reference. 
Two pulse spectra where the probe is co- (black) and cross- (red) polarized with the pump. Pump detuning = +0.83~meV and $\tau_{\mathrm{delay}}$ = 10~ps. The peak at a detuning of -1.96~meV corresponds to the $|X\rangle \to |2X\rangle$ transition. The insert shows the level diagram for the exciton-biexciton system, with allowed circularly polarized transitions.}
\label{fig:two_pulse_spec}
\end{figure}

To demonstrate the phonon-assisted population inversion, a two-pulse inversion recovery experiment \cite{Kolodka2007,Muller2013} was performed. First, the zero-phonon neutral exciton transition was found by measuring the absorption spectrum of a single $\pi$-pulse (blue line in Fig.~\ref{fig:two_pulse_spec}). Then a two-pulse experiment was performed with the pump pulse detuned by +0.83~meV, where the exciton population is predicted to be most efficiently created (see blue line in Fig. \ref{fig:detune_map}(a)), and the probe energy is scanned through the transition. The pump-pulse area is set to $8.46\pi$, the maximum available in our setup. The delay time between the pump and probe pulses $\tau_\text{delay}$ is 10 to 33.6 ps.

As illustrated in the energy-level diagram shown as the insert to Fig.~\ref{fig:two_pulse_spec},
the photon energy of the probe pulse and its polarization relative to the pump
selects the transition that is probed. We consider first the case for a
co-polarized probe pulse, where both the pump and probe pulses have the same
$\sigma^+$-polarization. In this case, the pump pulse excites a dephased, mixed
state exciton population. Since the probe has a
pulse area of $\pi$, it inverts the $z$-component of the Bloch-vector when it is resonant with the neutral exciton transition. The rotation of the state vector changes the exciton
population, resulting in a change in photocurrent proportional to the
populations after the pump but before the probe: $\Delta PC_{0-X}\propto C_0-C_X$. $\Delta PC$ is the change in the
photocurrent signal resulting from the dot that is induced by the probe
pulse. $\Delta PC$ is measured relative to the photocurrent measured for a detuned probe. A signature of
population inversion is that $\Delta PC_{0-X}$ should be negative, independent
of the photocurrent to exciton population calibration. Figure~\ref{fig:two_pulse_spec} shows the experimental results (black).  The  dip at zero-detuning clearly demonstrates that a population inversion has been
achieved between the $|0\rangle$ and $|X\rangle$-states.

The red line in Fig.~\ref{fig:two_pulse_spec} shows the results obtained for cross-polarized excitation ($\sigma ^+$ pump, $\sigma ^-$ probe). At zero-detuning, the probe addresses the orthogonally polarized 
exciton transition $|0\rangle\to|\bar{X}\rangle$, providing a measure of the occupation of the crystal ground-state $C_0$. The amplitude of the peak at zero detuning falls to less than half the amplitude measured by the single pulse, again confirming that a population inversion between $|0\rangle$ and $|X\rangle$ has been achieved since $\Delta PC_{0-\bar{X}} \propto C_0 - C_{\bar{X}}$. A second peak at a detuning of $-1.96~\mathrm{meV}$ corresponds to the $|X\rangle\to|2X\rangle$ transition, and provides a third measure of 
$C_X$ (see Ref.~\onlinecite{SMnote}).
\begin{figure}
\includegraphics[width=8.6cm]{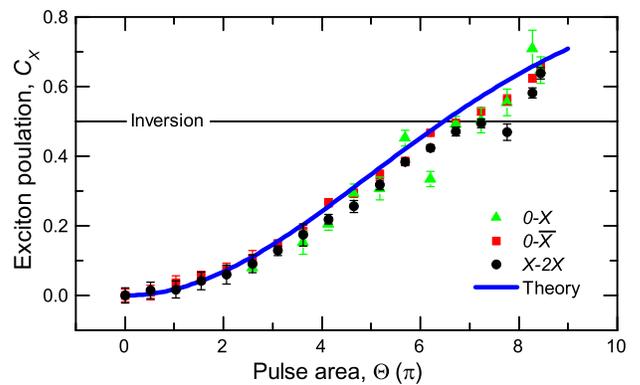}
\caption {Color online. The exciton population $C_X$ generated by the pump pulse with $\hbar \Delta = 0.83$~meV, as extracted from the exciton and biexciton peak in two-pulse spectra 
versus the pump pulse area. 
For a full derivation, see Ref.~\onlinecite{SMnote}. (Blue) Calculated $C_X$.}
\label{fig:amplitudes}
\end{figure}

Figure \ref{fig:amplitudes} plots the occupation of the neutral exciton versus
the area of the pump pulse. The three data points for each pulse area are
obtained from an analysis of the co and cross-polarized signals at zero
detuning, and from the biexciton peak. The exciton population is deduced by
comparison with the neutral exciton photocurrent peak for resonant excitation
with a single $\pi$-pulse (for details, see \cite{SMnote}). Since the electron
can tunnel out from the QD before the arrival of the probe, which reduces the
measured exciton population created by the pump, a correction is made to account
for an electron tunneling time of $\sim$50~ps. The three measurements of the
exciton population are in close agreement, and pass the transparency point,
$C_X=0.5$, at a pulse-area of $6.5- 7\pi$. The largest exciton population
observed is $0.67\pm 0.06$, limited by the power available in our setup. The
blue line in Fig.~\ref{fig:amplitudes} shows the results of 
our path-integral simulations which quantitatively reproduce the
experiments.

To investigate the dependence of the phonon-assisted population inversion on the pump frequency, a series of two-color photocurrent spectra, similar to Fig.~\ref{fig:two_pulse_spec}, were measured for a cross-polarized $7.24\pi$  pump pulse as a function of pump detuning with a $\pi$ probe. Figure~\ref{fig:detune_map}(a) presents the exciton population generated by the pump pulse. The spectrum has three features. At zero-detuning there is a pulsewidth-limited peak corresponding to the zero-phonon $|0\rangle\to|X\rangle$ transition. At positive detuning there is a broad feature due to  phonon emission. In principle, there can also be a phonon feature at negative detuning due to phonon absorption. However, the phonon absorption is negligible at low temperature. The third feature is a narrow peak at a pump detuning of $-1~\mathrm{meV}$, corresponding to the two-photon $|0\rangle \rightarrow |2X\rangle$ 
biexciton transition\cite{Stufler2006a}. This indicates that the pump pulse is slightly elliptically polarized.

\begin{figure*}
\includegraphics[width=17.8cm]{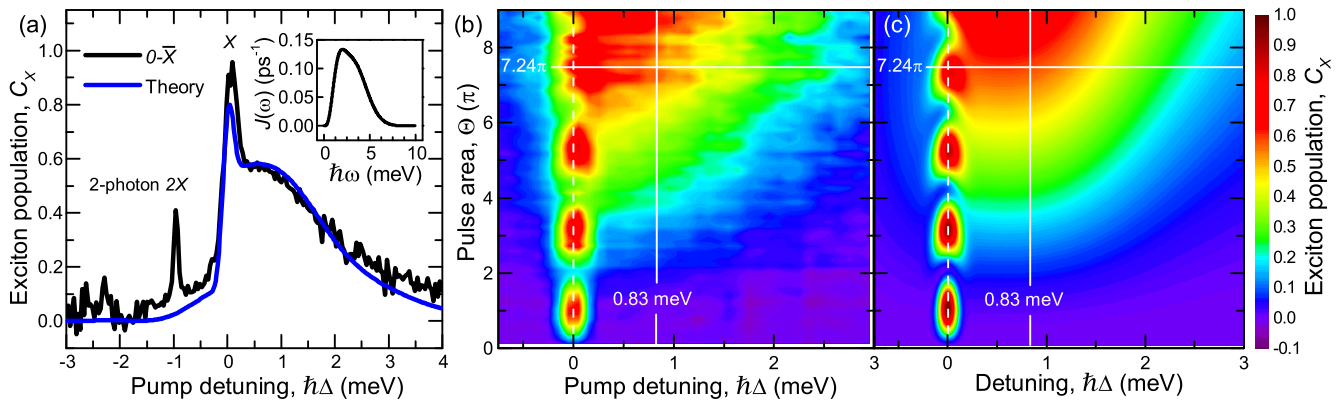}
\caption {Color online. (a) Exciton population created by a 7.24$\pi$ pump as a function of the pump detuning. $\tau_{\mathrm{delay}} =$~15~ps. Blue: Calculated $C_X$.  Insert: Theoretical spectral density of the exciton-phonon interaction $J(\omega$). (b) Experimentally obtained $C_X$ vs $\Theta$ and $\hbar\Delta$ with $\tau_{\mathrm{delay}} = 33.6$~ps. (c) Path-integral results for the same pulse area and detuning range as in (b). The vertical line is where $\hbar\Delta =$~+0.83~meV and horizontal line is for $\Theta = 7.24\pi$ as shown in figs.~\ref{fig:amplitudes} and Fig~\ref{fig:detune_map}(a), respectively.
}
\label{fig:detune_map}
\end{figure*}

In the theory, the phonon influence on the dot dynamics is mediated mainly by the
phonon spectral density: $J(\omega)=\sum_{\bf q} |\gamma_{\bf q}|^{2}\,\delta(\omega-\omega_{\bf q})$,
where $\gamma_{\bf q}$  is the exciton-phonon coupling.
In the absence of detailed information on the shape of the QD
we consider spherically  symmetric, parabolic potentials. 
For  bulk LA phonons coupled via the deformation potential we then obtain\cite{Vagov2011}:
\begin{multline}
J(\omega ) =\frac{\omega ^{3}}{4\pi ^{2}\rho \hbar v_{c}^{5}}
\left[
D_{\mathrm{e}}\,
e^{ \left( -\omega ^{2}a_{\mathrm{e}}^{2}/4v_{c}^{2} \right) }
-D_{\mathrm{h}}\,
e^{\left( -\omega ^{2}a_{\mathrm{h}}^{2}/4v_{c}^{2} \right) }
 \right]^{2},
\label{eq_J}
\end{multline}%
where $\rho$ is the mass density, $v_{c}$ the sound velocity and $D_{\mathrm{e/h}}$ denote the deformation potential
constants for electrons and holes. 
These material parameters are taken
from the literature  and are given explicitly in the supplement, while the electron and
hole confinement lengths $a_{\mathrm{e/h}}$ are used as fitting parameters. 
Good agreement with our experiments is obtained for
$a_{\mathrm{e}}=4.5$~nm and $a_{\mathrm{h}}=1.8$~nm resulting in the spectral density shown in the insert
of Fig.~\ref{fig:detune_map}(a). 
We note that the low frequency asymptote $\sim \omega^{3}$
is characteristic for bulk acoustic phonons and occurs independent of the material or the dot shape. On the other hand, the Gaussians in Eq.~(\ref{eq_J}) result from the Fourier transforms of the
electron and hole probability densities, reflecting the assumption of parabolic confinement potentials.

The blue line in Fig.~\ref{fig:detune_map}(a) shows the theoretical values of the
exciton population generated by the pump pulse, which excellently replicates  the
broadband feature observed at positive detuning. The lineshape of this feature is implicitly determined by $J(\omega)$, 
and more detailed information on $J(\omega)$ can be obtained than from fitting the intensity damping of Rabi rotations \cite{Ramsay2010}. 
Eq.~(\ref{eq_J}) shows that high frequency behavior of $J(\omega)$ follows from the Fourier transform of
the electron and hole probability densities, and so a lineshape analysis in principle provides a way to
learn something about the spatial distributions of the electrons and holes.
However, accessing this information would require  more detailed studies which are beyond the scope of the present letter. The asymmetry of the spectrum with respect to the sign of $\hbar\Delta$ unambiguously proves that the population created by the off-resonant pump pulse is the result of phonon-assisted relaxation into the
lower energy dressed state\cite{note:symmetric_sidebands}. We also note that the measured
low-energy phonon sideband is stronger than expected. This may be due to an
elevated temperature of about 6-7~K due to heating of the sample by the laser.

Figures \ref{fig:detune_map}(b) and (c) compare  the  exciton
population generated by the pump pulse measured as in Fig.~\ref{fig:detune_map}(a) at different pulse areas with corresponding path-integral calculations. On-resonance, the zero-phonon line exhibits intensity-damped Rabi
rotations. To positive detuning, there is the phonon-emission sideband that
broadens for higher pulse areas. 
The calculations
are in good agreement with the experimental data, further confirming the model. 
Even the slight shift of the resonant peaks towards higher energies with increasing pulse area is reproduced by the theory.

In conclusion, we have experimentally demonstrated the population inversion of a neutral exciton in a single quantum dot in a regime where the laser pulse is long compared to the time for the phonon-induced relaxation between photon-dressed dot states, 
following the theoretical proposal of Ref.~\cite{Glassl2013}. 
The application of a strong laser pulse at positive detuning increases the excitonic content of the higher energy ground-like dressed state, activating an LA-phonon emission process that causes relaxation to the lower-energy, exciton-like, dressed state. 
Here the phonon-bath provides added functionality to the quantum dot qubit, rather than simply acting as source of decoherence. Finally we note that, Madsen {\it et~al} \cite{Madsen2014} have proposed the use of phonon-assisted population inversion as a robust, non-resonant, deterministic pumping scheme for polarized single-photon sources.

This work was funded by the EPSRC (UK)
EP/J007544/1. MG, AMB and VMA gratefully acknowledge the financial support from Deutsche Forschungsgemeinschaft via the project AX 17/7-1. The authors thank H. Y. Liu and M. Hopkinson for sample growth.

\bibliography{Phonon_library}
\bibliographystyle{apsrev}

\section{Supplementary information}
\subsection{Extraction of the exciton population from measured data}
\label{population}

The exciton population created by a pump pulse tuned into the phonon sideband
can be extracted from our two-pulse spectra (see Fig.~2) in three different ways
by analyzing the peak heights of the three transitions: 
$|0\rangle \rightarrow |X\rangle$, $|0\rangle \rightarrow |\bar X\rangle$
and $|X\rangle \rightarrow |2X\rangle$. The relation
between the photocurrent ($PC$) signal and the exciton population is as follows. The $PC$  measured in our experiment is determined by the number of
electron-hole pairs in the sample, including pairs in the quantum dot (QD) and pairs
excited in the surrounding material by the pump and probe pulses, according to:
\begin{equation}
PC=\alpha(C^{'}_X+C^{'}_{\bar{X}})+2\beta C^{'}_{2X}+\gamma N^{'}_\text{s},
\label{Eq:PC}
\end{equation}
where $C^{'}_X$ and $C^{'}_{\bar{X}}$ are the populations of the two exciton states 
with opposite spins that are created in the QD 
after circularly polarized pump and probe pulses, while $C^{'}_{2X}$ is the biexciton population. 
$N^{'}_\text{s}$ is the number of electron-hole pairs created 
in the surrounding material which scales linearly with the laser intensity.
$\alpha$, $\beta$ and $\gamma$ are the detection efficiencies of our 
photocurrent measurement for each type of charge complexes. 
Specifically, for our pump-probe measurement, $PC$ can be further subdivided into two parts: 
a reference level $PC_\text{R}$ and the change of the photocurrent signal originating 
from the QD induced by the probe pulse $\Delta PC$ according to:

\begin{subequations}
\begin{align}
PC&=PC_\text{R} + \Delta PC,\\
PC_\text{R} &= (\alpha(C_X+C_{\bar{X}})+ 2\beta C_{2X})e^{-\tau_\text{delay}/T_1}
+ \gamma N^{'}_\text{s},\\
\Delta PC &= \alpha (\Delta C_X +\Delta C_{\bar{X}})+2\beta \Delta C_{2X},
\end{align}
\label{Eq:C}
\end{subequations}
where $C_{X,\bar{X},2X}$ are the exciton and biexciton
populations created immediately after the pump pulse
and $\Delta C_{X,\bar{X},2X}$ is the change of the exciton population induced
by the probe pulse in the QD. Since the electron can
tunnel out from the QD, the exciton and biexciton populations decay exponentially with
time. Thus, $C_{X,\bar{X},2X}e^{(-\tau_\text{delay}/T_1)}=C^{'}_{X,\bar{X},2X}$ represent the pump-created exciton or biexciton
populations that have remained in the QD  until the arrival of the probe pulse. 
Here, $\tau_\text{delay}$ is the delay time between the pump and probe
pulses and $T_1$ is the electron tunnelling time determined using inversion recovery measurement \cite{Kolodka2007}. In the above discussion, we have neglected 
the loss of the exciton population due to the radiative decay, since the radiative decay time
($\sim 600$~ps) is much longer than our typical delay times $\tau_\text{delay}$ (10 - 30ps). 
$PC_\text{R}$ is determined by measuring the photocurrent signal from 
an off-resonant measurement where  the pump pulse is tuned into the phonon 
sideband and the probe frequency is far from any of the resonances of the dot or the surrounding material
such that $\Delta PC$ becomes negligible and the photocurrent signal coincides with $PC_\text{R}$
in this case. 
$\Delta PC$ is the differential photocurrent signal that is discussed in the paper, 
which is determined from our experiment by subtracting the $PC_\text{R}$ obtained from the
off-resonant measurement from the total photocurrent signal $PC$. 
The detection efficiency $\alpha$ can be extracted from the single $\pi$ pulse experiment. 
Denoting by $\Delta PC(\pi)$ the maximum of the differential photocurrent signal reached 
with a single $\pi$ pulse we find:
\begin{equation}
\Delta PC(\pi)=\alpha C_X(\pi)=\alpha.
\label{Eq:alpha}
\end{equation}
Here, we have used that according to our path-integral simulations,
the phonon-induced deviation of $C_X(\pi)$ from the ideal value of 1 
is negligible at low temperatures.


Now let us derive the relation between the differential PC signal and the exciton population for 
the case when pump and probe pulse are co-circularly polarized and the probe pulse
is resonant to the $|0\rangle \rightarrow |X\rangle$  transition.
In this case, firstly, the $\sigma^{+}$ polarized pump pulse creates a certain exciton population $C_X$
and consequently the ground-state occupation after the pump pulse is given by $C_{0}=1-C_{X}$.
Then, in the time interval until the probe pulse arrives, the exciton population
is reduced to $C_Xe^{-\tau_\text{delay}/T_1}$ due to the tunneling. The ground-state occupation, on the other hand,
is not affected by the tunneling and thus stays at the value of $C_{0}=1-C_{X}$ until the arrival of the probe.
Finally, the $\pi$-pulse $\sigma^{+}$ polarized probe exchanges the populations of the
states $|0\rangle$ and $|X\rangle$ resulting in an exciton  population after the probe of $C^{'}_{X}=1-C_{X}$.
Therefore, the change of the $X$-exciton occupation induced by the probe pulse with the energy of $\hbar \omega_X$ 
is $\Delta C_X = 1-(1+ e^{-\tau_\text{delay}/T_1})\,C_{X}$. 
Using this result 
together with the fact, that for co-polarized $\sigma^{+}$-pulses the populations of the $\bar X$-exciton 
and the biexciton never build up, we find from Eq.~2(c):
\begin{equation}
\Delta PC_{0-X}=\alpha[1-(1+ e^{-\tau_\text{delay}/T_1})\,C_{X}]
\label{Eq:Hph}
\end{equation}
and thus with the help of Eq.~(\ref{Eq:alpha}) we eventually end up with:
\begin{equation}
C_X=\dfrac{1}{1+e^{-\tau_\text{delay}/T_1}}\left(1-\dfrac{\Delta PC_{0-X}}{\Delta PC(\pi)}\right).
\end{equation}

The situation is different when cross-circularly polarized pulses are used.  Let
us first discuss the case where the $\bar X$ polarized probe pulse is resonant
to the $|0\rangle \rightarrow |\bar X\rangle$  transition.  After the action of
the circularly $\sigma^{+}$ polarized pump pulse, the QD again has a certain probability $C_X$
to be in the $|X\rangle$ state and the probability to find the dot in the ground-state is $C_{0}=1-C_{X}$. The $\sigma^{-}$ polarized probe pulse induces transitions
from the ground-state to the $\bar X$ exciton. Since the probe pulse has a pulse
area of $\pi$ and the occupation of $|\bar X\rangle$ is zero before the arrival
of the probe, the probe pulse fully converts the occupation that was left in 
the ground-state after the pump pulse 
into an occupation of the $\bar X$ exciton, i.e., $C^{'}_{\bar X}=C_{0}=1-C_{X}$.
Again, the ground-state occupation is not affected by the electron tunnelling
and therefore no correction involving the tunnelling time $T_{1}$ should be applied.
Since the probe pulse is off-resonant to the $X-2X$ transition we can neglect the probe induced change
of the $|X\rangle$ and $|2X\rangle$ occupations.
Recalling that  $|\bar X\rangle$ is unoccupied before the probe, we find for
the resulting differential photocurrent signal $\Delta PC_{0-\bar{X}}$ with the 
probe pulse being in resonance to the exciton transition:
\begin{equation}
\Delta PC_{0-\bar{X}}=\alpha (1-C_X),
\label{Eq:Hph}
\end{equation}
which yields:
\begin{equation}
C_X=1-\Delta PC_{0-\bar{X}}/ \Delta PC(\pi).
\label{Eq:Hph}
\end{equation}

Besides from the data measured at the $|0\rangle \rightarrow |X\rangle$ 
and $|0\rangle \rightarrow |\bar X\rangle$ transitions, the exciton population
created by the pump can also be extracted from the exciton to biexciton
transition. An $\sigma^{+}$ polarized pump pulse tuned into the high-energy phonon
sideband of the neutral exciton transition again creates a certain exciton population
$C_X$ which evolves into $C_Xe^{-\tau_\text{delay}/T_1}$ until the arrival of the probe.
Biexcitons are not created, i.e., we have $C_{2X}=0$. A cross-polarized
$\pi$-power probe pulse resonant to the 
$|X\rangle \rightarrow |2X\rangle$ transition
converts the $|X\rangle$ population completely into
an $|2X\rangle$ population, which gives $C^{'}_X=0$ and
$C^{'}_{2X}=C_Xe^{-\tau_\text{delay}/T_1}$. Since the probe is now off-resonant to the
$|0\rangle \rightarrow |\bar X\rangle$ transition, the occupation of the
$|\bar X\rangle$ exciton induced by the probe is negligible 
and the ground-state occupation is not affected. 
Thus, the $\Delta PC$ signal resulting from a probe pulse 
in resonance to  the $|X\rangle \rightarrow |2X\rangle$
transition is given by:
\begin{equation}
\Delta PC_{X-2X}=2\beta \Delta C_{2X}+\alpha \Delta C_X=(2\beta - \alpha)C_Xe^{-\tau_\text{delay}/T_1}.
\label{Eq:PCbiexciton}
\end{equation}
$\beta$ can be determined from a separate experiment in whihc the pump is a $\pi$ pulse resonant with $X$ and the probe is a $\pi$ pulse resonant with $2X$. According to Eqs.~\eqref{Eq:alpha} and \eqref{Eq:PCbiexciton}, we have:
\begin{equation}
\beta=0.5(e^{\tau_\text{delay}/T_1}\Delta PC_{X-2X}(\pi)+\Delta PC(\pi)).
\label{Eq:beta}
\end{equation}

Inserting Eqs.~\eqref{Eq:alpha} and \eqref{Eq:beta} into Eq.~\eqref{Eq:PCbiexciton} 
we can extract the exciton population after the pump from:
\begin{equation}
C_X= \dfrac{\Delta PC_{X-2X}}{\Delta PC_{X-2X}(\pi)}.
\end{equation}

\subsection{Model}
\label{Model}

For our calculations we used the same model for an optically driven strongly 
confined quantum dot as in Ref.~\cite{Glassl2013},
which is based on the Hamiltonian
\begin{align}
 \label{eq:Hamiltonian}
 H = H_{\rm{QD-light}} + H_{\rm{QD-phonon}},
\end{align}
where  
\begin{align}
 H_{\rm{QD-light}} = \hbar\omega^{0}_{X}| X\rangle\langle X|
+\frac{\hbar\Omega(t)}{2} \left[ | 0\rangle\langle X| + |X\rangle \langle 0| \right],
\end{align}
and
\begin{align}
 H_{\rm{QD-phonon}} \!=\! \sum_{\bf q} \hbar\omega_{\bf q}\,b^\dag_{\bf q} b_{\bf q} 
\!+\! \sum_{\bf q} \hbar \big( \gamma_{\bf q} b_{\bf q} \!+\! \gamma^{\ast}_{\bf q} b^\dag_{\bf q}
                          \big) |X \rangle\langle X|.
\label{dot-ph}
\end{align}  
The ground-state $|0\rangle$ is chosen as the zero of the energy and
the phonon-free energy of the transition to the single exciton state $|X\rangle$  is denoted
by $\hbar\omega^{0}_{X}$.  The Rabi frequency $\Omega(t)$ is proportional to the
electric field envelope of a circularly polarized Gaussian laser pulse with
frequency $\omega_{L}$, which is detuned  from the ground-state to exciton
transition by $\Delta = \omega_{L}-\omega_{X}$, where $\omega_{X}$ is the
frequency of the single exciton resonance which deviates from $\omega^{0}_{X}$
by the polaron shift that results from the dot-phonon coupling
in Eq.~(\ref{dot-ph}). The coupling to the laser field
is treated in the common rotating wave and dipole approximations.  The operator
$b^\dag_{\bf q}$ creates a longitudinal acoustic (LA) bulk phonon with wave
vector $\bf{q}$ and energy $\hbar \omega_{\bf{q}}$.  We assume a linear
dispersion relation $\omega_{\bf{q}} = c_{s} |\bf{q}|$, where $c_{s}$ denotes
the speed of sound.  The phonons are coupled via the deformation potential only
to the exciton state. This coupling is expressed by the exciton-phonon coupling
$\gamma_{\bf{q}}=\frac{|\bf{q}|}{\sqrt{2V\rho \hbar \omega_{\bf{q}}}}
\left(D_{\rm{e}} \Psi^{\rm{e}}({\bf q}) - D_{\rm{h}} \Psi^{\rm{h}}({\bf
q})\right)$, where $\rho$ denotes the mass density of the crystal, $V$ the mode
volume, $D_{\rm{e/h}}$ the deformation potential constants, and
$\Psi^{\rm{e/h}}(\bf{q})$ the formfactors of electron and hole, respectively. As
explained in the main article, we calculate the formfactors from the
ground-state wavefunctions of a spherical symmetric, parabolic confinement
potential.  It should be noted that, in the pure dephasing model for the
dot-phonon coupling, no transitions between the bare electronic states can be
induced by the continuum of LA phonons, which can change the electronic
occupations only in the presence of the laser field. We assume the system to be
initially in a product state of a thermal phonon-distribution at the temperature
of the cryostat and a pure ground-state of the electronic subsystem.  We use the
material parameters given in Ref.~\cite{Krummheuer2002} for GaAs, which are: $\rho =
5370 \; \rm{kg}/\rm{m}^3$, $c_{s} = 5110 \; \rm{m}/\rm{s}$, $D_{\rm{e}} = 7.0 \;
\rm{eV}$, and $D_{\rm{h}} = -3.5 \; \rm{eV}$.

To obtain the time evolution of the electronic density matrix elements predicted 
by this model, we make use of a numerically exact real-time path-integral
approach, described in detail in Ref.~\cite{Vagov2011}.  This gives us the
opportunity to calculate the dynamics of the quantum dot with a high and
controllable numerical precision and without further approximations to the
given Hamiltonian. This includes taking into account all multi-phonon processes
and non-Markovian effects.

\end{document}